\documentclass[11pt]{article}
\pdfoutput=1
\usepackage{jheppub}

\usepackage[usenames,dvipsnames,table]{xcolor}
\usepackage{graphicx,amsmath,amssymb,amsthm,multirow,array,bm,bbm,esint}
\usepackage[mathscr]{eucal}
\usepackage[bbgreekl]{mathbbol}
\usepackage{epsf,amsfonts}
\usepackage{slashed}
\usepackage[numbers,sort&compress]{natbib}
\usepackage{minibox}
\usepackage{array,tikz-cd}
\usepackage{slashed}
\usepackage{ccaption}

\newcommand{\beq}{\begin{equation}}
\newcommand{\eeq}{\end{equation}}
\newcommand{\bea}{\begin{eqnarray}}
\newcommand{\eea}{\end{eqnarray}}

\newcommand{\prn}[1]{\left ( #1 \right )}
\newcommand{\brk}[1]{\left [ #1 \right ]}
\newcommand{\bigbr}[1]{\Bigl\{ #1 \Bigr\} }

\newcommand{\half}{\frac{1}{2}}

\newcommand{\Kbeta}{{\bm{\beta}}}
\newcommand{\LambdaB}{\Lambda_{\bm{\beta}}}

\newcommand{\Lag}{{\mathcal L}}

\newcommand{\skR}{\text{\tiny R}}
\newcommand{\skL}{\text{\tiny L}}

\newcommand{\UT}{U(1)_{\scriptstyle{\sf T}}}

\newcommand{\PS}{{\rm H}_S}

\newcommand{\LS}{{\overline{\rm H}}_S}

\newcommand{\delKMS}{{\Delta}_{_ {\bm\beta}} }

\newcommand{\Op}[1]{\mathbb{#1}}
\newcommand{\OpH}[1]{\widehat{\mathbb{#1}}}

\newcommand{\SKR}[1]{\mathbb{#1}_{\skR}}
\newcommand{\SKL}[1]{\mathbb{#1}_{\skL}}

\newcommand{\SKDif}[1]{\mathbb{#1}_{_{dif}}}

\newcommand{\SKAdv}[1]{\mathbb{#1}_{adv}}
\newcommand{\SKRet}[1]{\mathbb{#1}_{ret}}

\newcommand{\SKG}[1]{\mathbb{#1}_{_G}}
\newcommand{\SKGb}[1]{\mathbb{#1}_{_{\overline{G}}}}

\newcommand{\QSK}{\mathcal{Q}_{_{SK}}}
\newcommand{\QSKb}{\overline{\mathcal{Q}}_{_{SK}}}
\newcommand{\QKMS}{\mathcal{Q}_{_{KMS}}}
\newcommand{\QKMSb}{\overline{\mathcal{Q}}_{_{KMS}}}
\newcommand{\Q}{\mathcal{Q}}
\newcommand{\Qb}{\overline{\mathcal{Q}}}

\newcommand{\QC}{\Q}
\newcommand{\QCb}{\Qb}
\newcommand{\Qzero}{\Q_0}

\newcommand{\FSgn}[1]{(-1)^{F_{\mathbb{#1}}}}

\newcommand{\gradcomm}[2]{ \brk{ #1, #2 }_{\scriptscriptstyle \pm} }

\newcommand{\psib}{\overline{\psi}}

\newcommand{\rhoi}{\rho_{_0}}
\newcommand{\betai}{\beta_{_0}}
\newcommand{\diffBi}{\delta_{\Kbeta_{_0}}}

\title{The Fluid Manifesto: \\Emergent symmetries, hydrodynamics, and black holes}

\author[a]{Felix M. Haehl}
\author[b]{\!, R.\ Loganayagam}
\author[c]{\!, Mukund Rangamani}
\affiliation[\,a]{Centre for Particle Theory \& Department of Mathematical Sciences,\\
Durham University, South Road, Durham DH1 3LE, UK.}
\affiliation[\,b]{Institute for Advanced Study, Einstein Drive, Princeton, NJ 08540, USA.}
\affiliation[\,c]{Center for Quantum Mathematics and Physics (QMAP)  \\
Department of Physics, University of California, Davis, CA 95616 USA}
\emailAdd{f.m.haehl@gmail.com}
\emailAdd{nayagam@gmail.com}
\emailAdd{mukund@physics.ucdavis.edu}

\abstract{ We focus on  the question of  how relativistic fluid  dynamics should be thought of as a Wilsonian effective field theory emerging from Schwinger-Keldysh path integrals. Taking the basic principles of Schwinger-Keldysh formalism seriously, we are led to a series of remarkable statements and conjectures, which we phrase in terms of a broad programme relating relativistic fluid dynamics and  topological sigma models. Apart from the intrinsic interest for these ideas from the  non-equilibrium field theory viewpoint, we also emphasize its relevance to various  fundamental questions in black hole physics.\footnote{ This short note summarizes two talks at Strings 2015, Bengaluru, by two of the authors. }
}

\begin{document}
\maketitle


\section{Motivation}
\label{sec:intro}

Our broad interest is in understanding hydrodynamics as an effective field theory controlled by symmetries.  While the phenomenological aspects of hydrodynamics have been well understood for a long time, deriving them from a theoretical framework has been stymied by the absence of a clean framework to tackle integrating out degrees of freedom in  a mixed state  or density matrix of a quantum field theory (QFT).

The most general non-equilibrium dynamics of a microscopic QFT is thought to be described by
a Schwinger-Keldysh (complex-time) contour  formalism \cite{Schwinger:1960qe,Keldysh:1964ud}, where all microscopic fields and associated  symmetries are doubled and specific boundary conditions are imposed on the correlators of the two copies. On the other hand,  the fundamental assumption implicit in all of fluid dynamics is  that, in an appropriate macroscopic regime, this  doubled theory can be recast into dynamics of fluid fields (cf.,
\cite{landau} for a classic review).\footnote{ The basic axioms of hydrodynamics admit constructive solutions to arbitrary order in an effective field theory sense as proved recently in \cite{Haehl:2014zda,Haehl:2015pja}.}
 If the QFT has a classical gravitational description \cite{Maldacena:1997re}, via gauge/gravity duality the Schwinger-Keldysh theory can be recast into a third  gravitational description. When dealing with near-thermal states this gravitational description involves the dynamics of black holes \cite{Israel:1976ur,Maldacena:2001kr}.\footnote{ More generally we expect that the  field theory Schwinger-Keldysh contours to map into corresponding path integral contours for the gravitational theory, cf., \cite{Herzog:2002pc,Skenderis:2008dg} for developments of these ideas.}

However, the details of how this mapping or triality between these three descriptions works is ill-understood. For example, one may ask the  following questions:
\begin{enumerate}
\item The conventional fluid dynamics by itself does not  exhibit any doubling of fields. How does one reconcile this fact with the doubling  structure present in the microscopic Schwinger-Keldysh theory?
\item Fluid dynamics is crucially controlled by the presence of a local entropy current whose divergence is non-negative (second law of thermodynamics) \cite{landau}. How does this structure emerge from the doubled theory?
\item Via gauge/gravity duality, these questions translate into questions about black holes (for these are dual to fluid description of the gauge theory \cite{Bhattacharyya:2008jc})\footnote{ In particular, the fluid/gravity correspondence relates the hydrodynamic description of QFT plasmas to dynamical, inhomogeneous black hole solutions; see \cite{Rangamani:2009xk,Hubeny:2011hd} for reviews. } and  how they emerge out of an underlying  doubling of degrees of freedom (black hole complementarity).
\end{enumerate}
Motivated by these questions, we are  interested in developing a formalism which clarifies  the relation between these seemingly disparate descriptions.

The aim of this note is to conjecture a scenario which unifies these ideas, and outlines a programme which serves to derive the effective field theory of hydrodynamics. Our goal here is to present the physical arguments in favour of  a novel framework, eschewing the technical details, which will appear in  future publication(s).  Our primary thesis  in \S\ref{sec:micro} will be to argue that there is an efficient way to phrase the Schwinger-Keldysh construction in a manner amenable to coarse-graining RG transformations by invoking a set of topological symmetries inherent in the formalism. We will then point out that elements of this are already visible in recent constructions of hydrodynamic effective field theories in \S\ref{sec:macro}. Finally in \S\ref{sec:gravity} we speculate how these elements might conspire to address some of the vexatious issues in black hole physics. 

In keeping with our general philosophy of presenting  the broad physical picture, we have chosen to relegate an explicit construction illustrating these principles to Appendices. We urge the reader to consult them, as they serve as necessary supplementary material for the somewhat abstract discussion in the main text.  In Appendix \ref{sec:langevin} we explain our ideas in the simplest model of dissipative dynamics, viz., the Langevin system. Further conceptual issues relating to the entropy current are described in Appendix \ref{sec:utgauge}.

\medskip
\noindent
{\em Caveat lector:} In interest of full disclosure, we note that while the general principles we lay out appear to work as expected, we are yet to dot all the i's and cross all the t's, so there may yet be some nuances as to how certain technical aspects of the story actually play out.

\section{Microscopic story: Schwinger-Keldysh description}
\label{sec:micro}

We will begin with the Schwinger-Keldysh description \cite{Schwinger:1960qe,Keldysh:1964ud} (for a review see \cite{Chou:1984es}). The most general non-equilibrium dynamics of a microscopic QFT is thought to be described by a Schwinger-Keldysh (henceforth SK) path integral which evolves arbitrary mixed states of the field theory.  The SK path integral $\mathscr{Z}_{SK}$ is defined by the trace
\begin{equation}
\mathscr{Z}_{SK}[{\cal J}_\skR,{\cal J}_\skL] \equiv \text{Tr}\bigbr{\ U[{\cal J}_\skR]\ \rhoi\ U^\dag[{\cal J}_\skL]\ } \, ,
\label{eq:ZSKdef}
\end{equation}
where $\rhoi$ is the  initial density matrix, $U$ represents the unitary evolution operator of the QFT, $U^\dag$ is its adjoint and these two operators are deformed
by right and left sources ${\cal J}_\skR$ and ${\cal J}_\skL$ respectively. Functional differentiation with respect to these sources then computes correlators of the schematic form
\begin{equation}
\text{Tr}\bigbr{\ \rhoi \ \bar{\mathcal{T}} \prn{U^\dag \SKL{O} U^\dag \SKL{O}\ldots}\ \mathcal{T} \prn{U \SKR{O} U \SKR{O}\ldots}\ } \,,
\label{eq:SKCorr}
\end{equation}
where $\bar{\mathcal{T}}$ denotes anti-time ordering, $\mathcal{T}$ denotes time-ordering and we note that left operators are ordered to the \emph{left} of the right operators (thus justifying the terminology).

The SK description should be contrasted against the more familiar Feynman path-integral description of the QFT
\begin{equation}
\mathscr{Z}_{\text{Feynman}}[J] \equiv \langle \text{Vaccum}_{t=\infty} |\ U[{\cal J}]\  | \text{Vaccum}_{t=-\infty} \rangle \,,
\label{}
\end{equation}
which computes time-ordered correlators of the form\footnote{ A note on our convention: operators  of the original single copy microscopic theory are hatted, while
the doubled operators are denoted explicitly by appropriate subscripts. These  operators could be either elementary fields of the microscopic theory or more generally composite operators built from them. }
\begin{equation}
\langle \text{Vaccum}_{t=\infty} |\ \mathcal{T} \prn{U \OpH{O} U \OpH{O}\ldots}\  | \text{Vaccum}_{t=-\infty} \rangle \, .
\label{}
\end{equation}
Compared to this Feynman description, there are three central features of SK description as we will describe now.

\subsection{Doubling}

The first central feature  is the most well-known: {\em doubling} . In this formalism, all microscopic fields (and their symmetries) are doubled in order to represent both $U$ and $U^\dag$. This is natural from the perspective of working with a density matrix, which being an operator acting on the Hilbert space requires us to keep track of both the state and its conjugate.  We will refer to the two SK degrees of freedom as the right and the left copy of fields as presaged above. Consequentially we work in an enlarged Hilbert space ${\cal H}= {\cal H}_{\skR} \otimes {\cal H}_{\skL}$ with the action ($\Phi$ denoting the collection of fields)
\begin{equation}
S_{SK} = S[\Phi_\skR] - S[\Phi_\skL]\,.
\label{}
\end{equation}

It will be crucial in the sequel to note the relative sign between the two copies, which is predicated by the fact that while states are evolved forward (in the Schr\"odinger picture say), their conjugates are evolved in reverse under standard unitary Hamiltonian evolution. In particular, computing correlation functions involves turning on sources for the operators on both sides with a relative sign, or equivalently working with a Lorentzian inner-product in the source operator space, viz.,
\begin{equation}
\delta S_{SK} = \int d^dx \,\sqrt{-g}\left( {\cal J}_\skR \, \SKR{O} -{\cal J}_\skL \, \SKL{O} \right).
\label{eq:SKSources}
\end{equation}
This feature is manifest in the definition of the SK path integral $\mathscr{Z}_{SK}$ given in Eq.~\eqref{eq:ZSKdef}.

\subsection{Topological limit}

The second defining feature is a specific boundary condition imposed on the double copy correlators \cite{Chou:1984es}. Usually this is stated as a technical condition that  right-right correlators are all time-ordered, left-left correlators are all anti-time ordered and the left operators are always ordered to the left of the right operators; see  Eq.~\eqref{eq:SKCorr}. While  technically sufficient, this way of framing  is somewhat unwieldy to deal with. For example, it is not immediately clear how or why such an ordering structure should be preserved under renormalization.

We will thus rephrase this feature in a more useful form for doing effective theory. A consequence of the time-ordering prescription given above is that a certain class of operators, viz., the difference operators $\SKDif{O} = \SKR{O}  - \SKL{O}$, in the doubled theory have vanishing self-correlations.\footnote{ This statement is very familiar in the context of two point functions, where the advanced, retarded and the symmetric correlator form a complete basis. One can check that this statement extends trivially in the case of higher point functions, noting that it is a consequence of a simple identity involving time-ordering of operators \cite{Chou:1984es}. This identity is sometimes called the Veltman's largest time equation in the context of Cutkosky cutting rules \cite{tHooft:1973pz}.}  This is a manifestation of unitarity in the underlying QFT. In order to see this, we first note that, according to Eq.~\eqref{eq:SKSources}, difference operator correlators are computed by aligning the sources ${\cal J}_\skR={\cal J}_\skL= {\cal J}$. Looking at Eq.~\eqref{eq:ZSKdef}, it is clear that the SK path integral degenerates in this limit to a trace over initial state $\text{Tr}\left( \rhoi \right) $ if the evolution
with arbitrary sources is unitary.\footnote{ This shows that SK path integral is the right framework to study unitarity in the evolution of mixed states. This is to be contrasted with the thermofield double description which studies path integrals of the form $\text{Tr}\bigbr{\ U[{\cal J}_\skR]\ \rhoi^{\half}\ U^\dag[{\cal J}_\skL]\  \rhoi^{\half}}$ and is hence ill-suited for studying single copy unitarity unless it is analytically continued to a SK path integral. }

Typically it is hard to protect an entire set of correlation functions against correction without some symmetry principle. We therefore intuit there is underlying topological symmetry in play, since the above structure is insensitive to the particularities of the dynamics of the QFT under consideration.

Inspired by this observation, we take the second central feature of Schwinger-Keldysh description to be its cohomological/topological structure: {\em any Schwinger-Keldysh theory should be topological when considering only difference operators}.\footnote{ We can elevate the vanishing of difference operator correlations to a slightly stronger statement. Consider the case where ${\cal J}_\skR={\cal J}_\skL= {\cal J}$ after some time $t_1$. Then the SK path integral becomes topological for $t>t_1$ and hence localizes on the segment of the contour with $t_i < t < t_1$. This implies a precise notion of causality for SK correlators whose future-most insertions are the difference operators $\SKDif{O}$.}

Thus, doubling alone  is not sufficient -- the doubling should work in such a way that when the  sources for average operators  are turned off, the doubled theory should degenerate to a topological field theory. This topological feature is what makes the doubled theory equivalent to the original (undoubled) QFT and in particular ensures that the original unitarity is preserved. Since Witten's famous work on supersymmetric quantum mechanics \cite{Witten:1982im}, topological theories in higher dimensions are often identified as originating from twisted supersymmetric theories. So, this second feature can also be phrased as imposing a twisted supersymmetry in the SK framework.

More precisely, we elevate the doubling feature of Schwinger-Keldysh path integral into a quadruple.  In every SK  doubled theory, every operator $\OpH{O}$ is represented by a quadruplet  $\{\SKL{O},\SKGb{O},\SKG{O},\SKR{O}\}$. We will  assign the \textbf{Schwinger-Keldysh ghosts} $\SKGb{O}$ and $\SKG{O}$ opposite Grassmann parity (but the same fermionic parity and hence the same  thermal boundary conditions) as the original operator $\OpH{O}$. We  will also demand the existence of \textbf{Schwinger-Keldysh supercharges} $\QSK$ and $\QSKb$ which are mutually anti-commuting, Grassmann odd, nilpotent operators with zero fermion number. They are defined by the graded commutators:
\begin{equation}\label{eq:QSKdefRL}
\begin{split}
\gradcomm{\QSK}{\SKL{O}}& = \gradcomm{\QSK}{\SKR{O}}  = \SKG{O},\quad
\gradcomm{\QSK}{\SKG{O} } = 0 ,\quad
\gradcomm{\QSK}{\SKGb{O}} = -\prn{\SKR{O}-\SKL{O}} . \\
\gradcomm{\QSKb}{\SKL{O}}& = \gradcomm{\QSKb}{\SKR{O}}  = \SKGb{O},\quad
\gradcomm{\QSKb}{\SKGb{O} } = 0 ,\quad
\gradcomm{\QSKb}{\SKG{O}} = \prn{\SKR{O}-\SKL{O}} .
\end{split}
\end{equation}
We resort to a simple notation where  $\gradcomm{A}{B}$ denotes a commutator if either $A$ or $B$ is Grassmann even and an anticommutator otherwise. This can be represented diagrammatically as:
\begin{equation}
\begin{tikzcd}
&\SKR{O}, \SKL{O} \arrow{ld}{\QSK} \arrow{rd}[below]{\QSKb}   &   \\
\SKG{O}\arrow{rd}{\QSKb} & & \SKGb{O} \arrow{ld}[above]{\!\!\!\!\!\!\!\!\!\!\!\!\!\!-\QSK}\\
&   \SKR{O}- \SKL{O} &
\end{tikzcd}
\label{eq:qskaction}
\end{equation}

The presence of these supercharges then defines and controls the topological subsector of difference operators in any SK theory. The SK ghosts with their attendant nilpotent
symmetries thus ensure microscopic unitarity (very much like how Faddeev-Popov ghosts with nilpotent BRST invariance ensure unitarity in gauge theories).

Let us now examine this topological limit more closely. As described above when the sources are aligned the SK path integral becomes topological in that it only depends on the boundary data, viz., on the initial mixed state:
$$
 \mathscr{Z}_{SK}[{\cal J}_\skR={\cal J}_\skL] =\text{Tr}\left( \rhoi\right) \,.
$$
If the initial density matrix is normalized (which is equivalent to subtracting out the topological contributions), the SK path integral just becomes unity in this limit. It is however
convenient to \emph{not} normalize the initial density matrix. For example, we will be interested in thermal initial conditions with the un-normalized density matrix $\rhoi = e^{-\betai\OpH{H}}$, where $\OpH{H}$ is the Hamiltonian which dictates the dynamics of the theory. Such a state admits a ready Euclidean path integral representation. In that case, we have a clear defining requirement for the topological field theory under question:  it should localize onto an appropriate thermal path integral (or partition function) over its initial time boundary:
\begin{equation}
    \mathscr{Z}_{SK}[{\cal J}_\skR={\cal J}_\skL] =\text{Tr}\left( e^{-\betai\OpH{H}} \right)= 
    \mathscr{Z}_{Euc}[\betai] \,.
\label{eq:skEth}
\end{equation}
At least for (near) thermal systems, one can thus heuristically think of the topological subsector as encoding the relevant entanglement structure for the class of
near thermal mixed states being studied.

We  interpret Eq.~\eqref{eq:skEth} as follows: $\mathscr{Z}_{Euc}$ encodes the  equal-time correlation functions of the initial thermal state and hence the
 fundamental entanglement structure necessary for SK path integral. On the other hand, $\mathscr{Z}_{SK}[{\cal J}_\skR={\cal J}_\skL]$  encodes the correlations of the difference operators. Since unitary evolution away from this initial state involves forward evolution on the right and backward evolution on the left, the
 subsector of difference operators is sensitive solely to the initial correlations encoded in $\mathscr{Z}_{Euc}$, irrespective of where they are inserted in time.
 Thus, $\mathscr{Z}_{SK}[{\cal J}_\skR={\cal J}_\skL]$ should be a topological field theory of initial correlations/entanglement structure which localises to $\mathscr{Z}_{Euc}$, leading thus to \eqref{eq:skEth}.

 The above interpretation  leads to a clear strategy to do effective theory of SK path integrals. One first writes down a ``{\em topological backbone}" theory that captures appropriate correlations/entanglement of the initial mixed state under study. Then we deform away from the ${\cal J}_\skR={\cal J}_\skL$ limit to study the class of mixed states which are continuous deformations of this initial state. These are mixed states which have similar entanglement structure (captured by a single topological field theory)
 and which evolve into each other under unitary evolution.\footnote{ Another way of phrasing this argument in a language sympathetic to the quantum information perspective is the following: the SK construction employs a topological invariance to first set-up a sector with built in robustness against non-topological perturbations. This topological backbone can then  heuristically be thought of as a continuum equivalent of building a tensor network for the initial state. Indeed the construction above is reminiscent of tensor networks employed for constructing thermal density matrices in spin systems \cite{Evenbly:2014aa}.   This analogy stems from the fact that in both methods we first  encode the relevant entanglement and then think of  dynamics as a continuous deformation  keeping the basic entanglement structure intact. } This strategy of imposing topological invariance is clearly  a natural thing to do from an effective field theory viewpoint and, being a symmetry, it is robust
 against renormalization.

We will end this discussion with a remark aimed at making contact with existing literature: first  the SK superalgebra is closely related to field redefinitions in the original theory which are viewed as being gauge fixed by the equation of motion (or more precisely Schwinger-Dyson equations) of the theory \cite{Alfaro:1992np}. In the topological field theory literature, this is viewed as constructing topological field theory by `gauge fixing a  zero action theory'  \cite{Birmingham:1991ty}.

\subsection{The KMS condition}

Next we turn to the third  and last defining feature of Schwinger-Keldysh formalism. Unlike the first two aspects discussed above, this third set of properties are peculiar to thermal states (and presumably near-thermal states like fluid states).  The conventional way to state this is to impose them at the level of correlators in the form of {\em Kubo-Martin-Schwinger (KMS) conditions} \cite{Kubo:1957mj,Martin:1959jp}, which assert that the thermal equilibrium correlators are periodic in Euclidean time. These are non-local\footnote{ The canonical scale of non-locality is of course just the thermal scale.} conditions on thermal Schwinger-Keldysh correlators which ensure that they are related to Euclidean correlators by analytic continuation.

Consider the situation where the initial state is an equilibrium thermal state. On the initial time slice, we can then formulate the thermal (KMS) boundary condition  imposing Euclidean periodicity as
\begin{equation}\label{eq:KMSbasic}
\FSgn{O} e^{-i\diffBi} \OpH{O}\prn{t=t_i}= \OpH{O}\prn{t=t_i} \,.
\end{equation}
The operator $\FSgn{O}$ imposes periodic or anti-periodic boundary conditions along imaginary time, depending on whether the observable is bosonic or fermionic.\footnote{ Thus $F_\Op{O}$ is the fermion number operator of the field theory in question. It is taken to be zero for Grassmann even particles and  Grassmann odd ghosts both of which we refer to as `bosonic'. $F_\Op{O}$ is unity for Grassmann odd particles and  Grassmann even ghosts both of which we refer to as `fermionic'. In this terminology, all bosonic fields are given periodic boundary conditions and all fermionic fields are given anti-periodic boundary conditions along thermal circle. We note that this ensures that BRST invariances which relate say bosonic particles to bosonic ghosts are preserved by thermal boundary conditions.  \label{fn:bcs} }  On the other hand $\diffBi$  is a classical differential operator. Physically, one should view it as  enforcing  a thermal  Lie-drag along the timelike vector $\Kbeta_0^\mu$. The latter is the background Killing vector whose presence is guaranteed by stationarity of the initial state (see \cite{Haehl:2015pja} for a definition in hydrodynamics). 

It is perhaps useful at this juncture to clarify the following: it is for the most part common to assume that one is discussing thermal physics of a QFT in flat space, whence $\diffBi$ corresponds simply to a thermal shift along the Euclidean circle as in the usual Hamiltonian time evolution. Formally, we may write (for scalar operators) $\diffBi = \betai\, \frac{d}{dt}$, so that we may infer: 
\begin{equation}\label{eq:BetaShift}
 e^{-i \betai  \frac{d}{dt}} \, \OpH{O}(t) \equiv e^{-\betai \OpH{H}} \; \OpH{O}(t) \, e^{\betai \OpH{H}} =  \OpH{O}(t-i\betai) \,.
\end{equation}
When considering curved manifolds with a timelike Killing field it is more appropriate to think of $\diffBi$ as implementing a Lie drag in the direction along the local Euclidean circle (whose size may vary spatially).  Furthermore, we should emphasize that $\diffBi$ should not be viewed simply as a diffeomorphism in the Euclidean time direction. Since this is a feature of the Gibbs density matrix, the best way to describe it is in terms of a state-dependent (thermal) time translation.

The usual Wilsonian intuition about decoupling of scales implies that these nonlocal relations can be treated  as local relations at length scales  larger than thermal scales. Thus we  expect that there is a generalization of KMS relations to  near-thermal situations like fluid dynamics, where they appear in a local avatar. At present, how this happens
is not completely understood, but we will  have more to say regarding this issue when we discuss fluid dynamics.

For now, let us examine how KMS conditions appear in the Schwinger-Keldysh correlators. If we take the initial state to be exactly thermal, we can introduce a set of KMS conjugate operators  
\begin{equation}
\SKL{\tilde{O}}\equiv \FSgn{O}  e^{-i\diffBi} \SKL{O} \,,
\label{eq:kmsconj}
\end{equation}	
which are thermal time-translates of the operator in question. 
We could likewise also introduce in a similar fashion the analytically continued sources, 
\begin{align}
\tilde{{\cal J}}_\skL = \FSgn{\cal J} e^{-i\diffBi} {\cal J}_\skL \,,
\end{align}
and  similarly define analogous conjugations  for the Grassmann odd counterparts (see footnote \ref{fn:bcs}). In equilibrium the KMS condition guarantees us that we can replace $\{\SKL{O}, {\cal J}_L\} \;\to\; \{ \SKL{\tilde{O}}, \tilde{{\cal J}}_\skL \}$ and the physical correlation functions remain invariant. 
 
A corollary of the statement above is that arbitrary correlation functions of the operators $\SKR{O} -\SKL{\tilde{O}}$ vanish in the thermal state. This is usually stated as a sum rule \cite{Weldon:2005nr} and is independent from the one that guarantees the vanishing of difference operator correlators in the SK construction. From the perspective we advocate, it would be natural to associate a second topological structure associated with this set of vanishing correlation functions.

A natural consequence of this observation is that we can now repeat our arguments in the previous subsection with ${\cal J}_\skL$ replaced by  $\tilde{{\cal J}}_\skL$ and $\SKL{O}$ replaced by $\SKL{\tilde{O}}$. We then are led to conjecture that these local versions of generalized KMS relations relevant in fluid regime are completely fixed by a (second) dual topological structure. More precisely, consider  another set of mutually anti-commuting, Grassmann odd, nilpotent operators  called the \textbf{KMS supercharges} $\QKMS$ and $\QKMSb$  whose action is given as: 
\begin{equation}\label{eq:QKMSdefRL}
\begin{split}
\gradcomm{\QKMS}{\SKL{O}} = \SKG{O}\,,\qquad&
\gradcomm{\QKMS}{\SKR{O}} = \FSgn{O}e^{-i\diffBi}\SKG{O}\, ,\\
\gradcomm{\QKMS}{\SKG{O}} = 0\, ,\qquad&
\gradcomm{\QKMS}{\SKGb{O}} = -\left(\SKR{O}-\FSgn{O}  e^{-i\diffBi} \SKL{O}\right)\, ,\\
\gradcomm{\QKMSb}{\SKL{O}} =\SKGb{O}\,,\qquad&
\gradcomm{\QKMSb}{\SKR{O}} = \FSgn{O}e^{-i\diffBi}\SKGb{O}\, ,\\
\gradcomm{\QKMSb}{\SKGb{O}} = 0 \, ,\qquad&
\gradcomm{\QKMSb}{\SKG{O}} = \SKR{O}-\FSgn{O}  e^{-i\diffBi} \SKL{O}\, ,
\end{split}
\end{equation}
or in diagrammatic notation (which makes explicit the similarity to Eq.~\eqref{eq:qskaction}):
\begin{equation}
\begin{aligned}
\begin{tikzcd}
&\quad\SKL{O}\quad \arrow{ld}{\!\!\!\QKMS} \arrow{rd}[below]{\!\!\!\!\QKMSb}   &   \\
\SKG{O}\arrow{rd}{\!\!\!\QKMSb \quad\;-\QKMS} & & \SKGb{O} \arrow{ld}\\
&   \SKR{O}- \SKL{\tilde{O}} &
\end{tikzcd}
\end{aligned}
\label{eq:qkmsaction}
\end{equation}
and similarly for $\SKR{O}$.\footnote{  One can for instance check that $\FSgn{O}e^{i\diffBi}\SKR{O}$ has the same diagram as \eqref{eq:qkmsaction}. This can be derived  from a ${\mathbb Z}_2$ automorphism of the SK algebra, or more simply by noting that the SK contour should be traversed in the opposite orientation to go from $\text{L} \to \text{R}$.} One can understand the relations here quite straightforwardly by replacing the difference operators of the SK construction $\SKR{O} - \SKL{O}$ by their thermal counterparts $\SKR{O}-\SKL{\tilde{O}}$ which are engineered to be aware of the KMS condition.

It is clear that these supercharges do not act locally because of  thermal translations $e^{-i\diffBi}$, which relate fields separated along the (Euclidean) thermal circle. Thus, we expect a strict definition of these supercharges on a general state is probably  fraught with usual subtleties familiar in QFT. But, in the exact thermal state,  working in Fourier modes, these operators can be defined precisely
(by physicists' standards).\footnote{ Such non-local supercharges  appear in other contexts as well, cf., \cite{Cecotti:2010qn}.}   We should however note that $\QKMS$ commutes with $\OpH{H}$, since we are trying to encode the invariance under Euclidean evolution by a thermal period.

The KMS conditions then translate into imposing a dual cohomological structure associated with these supercharges such that  
\begin{equation}
 \gradcomm{\QSK}{\QKMSb} =- \gradcomm{\QSKb}{\QKMS} = 1 - (-1)^F \, e^{-i\diffBi} 
\end{equation}
and all other anti-commutators of the four supercharges $\{\QSK, \QSKb,\QKMS,\QKMSb\}$vanishing.  This can be easily checked given our definitions of their action in \eqref{eq:QSKdefRL} and \eqref{eq:QKMSdefRL},  coupled with the fact that the charges  commute with the Hamiltonian.

It is useful to record here that in thermal equilibrium we have KMS invariance. States respecting this condition  should be viewed as stationary states; the system admits a background timelike Killing field (which can be identified with $\Kbeta^\mu_0$). One consequence of KMS invariance is that the SK construction in the absence of sources for average operators (i.e., in the topological sector) collapses onto a single copy Euclidean theory which is thermal.  In the near-thermal regime allowing states with spatial variations implies that one encounters configurations relevant for hydrostatic equilibrium, a fact that will prove useful in the sequel.

\paragraph{Taking stock:} To summarize, we are arguing that the basic elements of the SK path integrals for real-time dynamics can be distilled into three essential features: (i) doubling degrees of freedom and symmetries, (ii) presence of a pair of field redefinition BRST symmetries $\{\QSK, \QSKb\}$, and (iii) an emergent pair of topological charges $\{\QKMS, \QKMSb\}$ when we restrict to near-equilibrium dynamics.

\subsection{A mathematical interlude} 
\label{sec:math}

 One can in fact make a much stronger statement.\footnote{ This section lies outside the main line of development of the paper and is presented here for completeness. Interested readers will find it helpful to consult the original source \cite{Dijkgraaf:1996tz} and Appendix \S\ref{sec:LangevinCartan} for an explanation of the statements described here. } Let us define yet another generator $\Qzero$ whose action on the quadruplet of fields $\{\SKR{O}, \SKL{O}, \SKG{O}, \SKGb{O}\}$ can be defined as\footnote{ Note that the $\Qzero$ action is very simple in a basis that is known in the thermal QFT literature as the advanced-retarded basis, where $\SKAdv{O} \equiv \SKR{O}  - \SKL{O}$ and 
 $\SKRet{O} \equiv -i \delKMS^{-1} \, (\SKR{O} - \SKL{\tilde{O}} ) $. Then $\gradcomm{\Qzero}{\SKRet{O}}=0$ and $\gradcomm{\Qzero}{\SKAdv{O}} = -i\delKMS\SKRet{O} $ This is clearly adapted to thermal density matrix owing to the explicit occurrences of the thermal translation operator in the definitions. See Appendix \ref{sec:langevin} where we work out the example of a Langevin particle in this advanced-retarded basis.
 \label{fn:advret}}
 \begin{equation}
 \begin{split}
   \gradcomm{\Qzero}{\SKL{O}} = \frac{1}{1-\FSgn{O} e^{-i\diffBi}} \left( \SKR{O} - \SKL{\tilde{O}} \right) \,,\qquad&
   \gradcomm{\Qzero}{\SKR{O}} = \frac{\FSgn{O} e^{-i\diffBi}}{1-\FSgn{O} e^{-i\diffBi}} \left( \SKR{O} -\SKL{\tilde{O}}\right)\,, \\
   \gradcomm{\Qzero}{\SKG{O}} = 0 \,,\qquad & \gradcomm{\Qzero}{\SKGb{O}} = 0 \,.
 \end{split}
 \end{equation}
We claim that the operator $i\delKMS \equiv 1 - (-1)^F \, e^{-i\diffBi} $ together with the generators thus far introduced $\{\QSK, \QSKb, \QKMS, \QKMSb, \Qzero\}$ comprise the generators of  the so called ${\cal N}_T = 2 $ extended equivariant cohomology superalgebra \cite{Dijkgraaf:1996tz}. 

These algebras which were first encountered in \cite{Vafa:1994tf} in the context of topologically twisted ${\cal N}= 4$ SYM, were realized to underlie twisted supersymmetric theories with two distinct pairs of supercharges, which have been christened balanced topological field theories. The natural interpretation in equivariant cohomology language is that $\{\QSK, \QSKb\}$ are the basic differentials, while $\{\QKMS, \QKMSb, \Qzero\}$ are the interior contractions and $\delKMS$ is the Lie derivative operator. To be specific one can work with the (extended) Weil model of equivariant cohomology -- the supercharges $\{\QSK, \QSK\} $ are then the Weil differentials, while the interior contractions are invoked to define the basic forms to ensure that the cohomology is non-trivial. It may further be of interest to the reader to note that the algebraic structure of these charges is essentially unique \cite{Zucchini:1998rz}. 

We recall here that the notion of equivariance reflects the existence of a group action that commutes across homomorphisms between two vector spaces. The group action we are invoking here is the abelian group generated by thermal time translations. The generators $\{\QKMS, \QKMSb\}$ precisely implement this particular translation and will be of great import in the hydrodynamic discussion we present below.

The above statements refer purely to the algebraic structure, eschewing for the moment the inherent complications in defining non-local charges. However, as we shall argue below, in the hydrodynamic limit these charges become local and ensure thus that we have a rigid algebraic structure underlying the class of effective field theories we seek to understand. A clean physical example which illustrates the basic features alluded to above is a Brownian particle undergoing Langevin dynamics, which we explain briefly in 
Appendix \ref{sec:langevin} (a complete discussion employing the equivariant language will appear elsewhere \cite{Haehl:2015ab}).

\section{Macroscopic story: Hydrodynamic description}
\label{sec:macro}

We now turn to how the microscopic physics described in the previous section leads to fluid dynamics in macroscopic scales. Explaining the emergence of fluid dynamics is a challenge in two fronts:  on one hand, we  have to explain what is the macroscopic fate of the three central features of the microscopic SK description. On the other hand, there are genuinely novel features of the fluid description, like the entropy current, whose microscopic origins need to be clarified.

\subsection{Schwinger-Keldysh features in hydrodynamics}

Let us begin by explaining what happens to the three features of the last section. Firstly, we propose that the doubled symmetries are spontaneously broken at macroscopic scales. More precisely, we take the difference symmetries to be one which are broken, and  therefore, giving rise to Goldstone modes. The claim here is that these Goldstone modes are the low energy fluid modes and the sigma model describing them is fluid dynamics. 

The symmetries which are broken correspond to the difference diffeomorphisms and difference flavour symmetries (for charged fluids). Consequentially, this implies that we have a vector Goldstone modes (for broken diffeomorphisms) and flavour Lie algebra valued scalar Goldstones (for broken flavour symmetries). This is indeed the correct count of degrees of freedom visible in hydrodynamics, as can be inferred for example from our recent discussion in \cite{Haehl:2015pja}, where we use an un-normalized vector field $\Kbeta^\mu = u^\mu/T$ and a Lie algebra valued flavour field $\LambdaB$.\footnote{ This set of variables as explained in \cite{Haehl:2015pja} has many advantages to the more traditional presentation, where the variables are taken to be a normalized (timelike) velocity field $u^\mu$, and the intensive parameters characterizing the density matrix, viz., temperature, chemical potential etc..}

Per se, this then resolves the question of why the doubling is hidden in the fluid description: it is because it is spontaneously broken and \emph{the very existence of fluid Goldstone modes is the macroscopic manifestation of the microscopic doubling}.

We next come to the second feature of SK path integrals: accounting for the fact  that the difference operator subsector is a topological field theory. This immediately implies that the sigma model under question should become topological in an appropriate limit. This can be achieved by demanding that the sigma model describing fluid dynamics  be a slight deformation of a \emph{topological} sigma model. This is synonymous with demanding that  the fluid dynamics is a \emph{twisted supersymmetric} sigma model with difference operators forming its cohomological  observables. We elaborate on this below.

Finally, we turn to the third and the most non-trivial feature of SK description: KMS relations and their generalizations. In the last section, we stated them in terms of a pair of non-local supercharges $\{\QKMS,\QKMSb\}$ acting via thermal translations and their cohomology. At long distances compared to the thermal scale, this set of constraints should become local. We propose that it turns into an abelian gauge invariance, which we denote by $\UT$ following our earlier work on hydrodynamic effective field theories \cite{Haehl:2014zda,Haehl:2015pja}.
We claim that this abelian  gauge invariance emerges on macroscopic scales in order to impose KMS invariance.  More specifically, the topological charges  $\{\QKMS,\QKMSb\}$ should be viewed as the Faddeev-Popov BRST and anti-BRST charges for $\UT$. Consequentially, fluid dynamics is a slight deformation of a \emph{gauged} topological sigma model.

As presaged in \S\ref{sec:math} is basically only one topological gauge theory with two topological charges; any such gauge theory has the structure of Vafa-Witten twisted $\mathcal{N}=4_{4d}$ SYM theory \cite{Vafa:1994tf}. More precisely, one has two topological charges, lets call them $\{{\cal Q}, {\bar {\cal Q}}\}$ which basically implement  two copies of the equivariant cohomology of the $\UT$ gauge group under question. Such topological field theories have been christened balanced topological field theories in \cite{Dijkgraaf:1996tz}. What we want is basically an $N_T =2$ cohomological field theory in their language, with the two twisted supercharges. One identifies the supercharges $\{{\cal Q}, {\bar {\cal Q}}\}$ as the  Cartan charges of the equivariant cohomology construction, which may be viewed in a suitable sense as the gauged fixed form of the SK supercharges  $\{\QSK,\QSKb\}$.  The $\UT$ supercharges which behave as the Faddeev-Popov BRST charges of the emergent Abelian symmetry group should be viewed as the gauge transformations of the equivariance group. The interested reader can see this structure quite readily in the context of a linear dissipative system such as Langevin dynamics which we review in Appendix \ref{sec:langevin}.

In the topological limit, i.e., when average operators are not sourced, the SK path integral equivariantly localizes on states which are invariant under the topological $\UT$ symmetry. Since $\UT$ acts by thermal translations, as described earlier  the KMS invariant states are precisely those which can be described by a Euclidean path integral, where the thermal direction has no local structure. We conclude that in this limit the SK path integral reduces to the Euclidean equilibrium partition function.\footnote{ This way of understanding hydrostatics as an endpoint of equivariant localization also gives a natural explanation for the appearance of `hatted' connections in the analysis of hydrostatic anomalies \cite{Jensen:2013kka,Haehl:2013hoa}. The hatted connections are connections shifted by a chemical potential, viz., $\hat{A}_\sigma = A_\sigma + \mu u_\sigma$ and are natural connections from equivariant cohomology viewpoint.} This also shows that SK path integrals are analytic continuations of Euclidean path integrals. Equivalently, the fluid-dynamic path integrals are analytic continuations of  hydrostatic path integrals of  \cite{Banerjee:2012iz,Jensen:2012jh}, where the latter can be formulated on a time slice, i.e., in one dimension lower.\footnote{  This is rather reminiscent of how the analytic continuation of Chern-Simons theory
\cite{Witten:2010cx} leads to the alternate twist of $\mathcal{N}=4_{4d}$ SYM
\cite{Marcus:1995mq,Kapustin:2006pk}. Both the Chern-Simons theory and hydrostatics live in one lower dimension and both of them have no propagating degrees of freedom. In both cases their analytic continuations live in one dimension higher and exhibit twisted supersymmetry which exploits equivariant localization for analytic continuation. This provides a heuristic motivation the origin of a Vafa-Witten balanced topological field theory in the hydrodynamic context.}

This proposal then comes with a bonus feature: we conjecture that  \emph{the abelian $\UT$ current is precisely the entropy current}. Thus, the presence of the entropy current is traced back to the KMS invariance in the microscopic description. This set of ideas taken together then explain the emergence of fluid dynamics.
In the rest of the section, we will list the evidence in favor of this proposal. 

\subsection{Evidence for a topological sector in hydrodynamics}

First of all, we can try writing down a topological sigma model for fluid dynamics by applying Mathai-Quillen formalism \cite{Mathai:1986tc}, or equivalently, gauged Martin-Siggia-Rose (aka MSR) formalism \cite{Martin:1973zz} on the fluid equations.\footnote{ See Appendix \ref{sec:langevin} for an explicit version of this construction in the case of the Langevin system. This serves as a one-dimensional toy model for the hydrodynamic field theory.} 

The basic idea of these formalisms is to construct an effective action that localizes on a solution of the classical field equations by introducing new auxiliary  degrees of freedom. In order to keep the count of the physical phase space correct these new variables are ghost fields leading to an topological field theory. Heuristically, in a nutshell, one simply introduces these auxiliary variables by examining the Faddeev-Popov identity which asserts that 
\begin{equation}
1 = \int [{\cal D}\Phi] \, \delta(\text{Eom}(\Phi)) \, \text{det}\left(\frac{\delta \text{Eom}(\Phi)}{\delta \Phi}\right)
\label{eq:fpidentity}
\end{equation}	
The determinant naturally gives rise to the ghost fields, while the exponential of the delta functional gives rise to the Lagrange multiplier field (the  Nakanishi-Lautrup field). Note that by virtue of the construction we naturally encounter the quadrupling indicated in the SK construction. The fields, the BRST ghost, anti-ghost and the Nakanishi-Lautrup field fill out a multiplet in a canonical manner.

 In the hydrodynamic context the equations of motion are simply energy-momentum and charge conservation equations along with the entropy production equation. One non-trivial observation we can make is that the Mathai-Quillen action of fluid equations naturally exhibits an abelian $\UT$  gauge invariance whose gauge field corresponds to the Nakanishi-Lautrup degree of freedom associated with entropy conservation \cite{Haehl:2015aa}.  As a result the entropy current along with topological invariance ala Mathai-Quillen implies $\UT$  gauge invariance, which then needs  to be fixed to make sense out of  the path integral. This is not a derivation of our proposal  per se, since we have assumed some parts of it in deriving some other part, but it is a non-trivial check of its consistency.

A further piece of evidence for our proposal stems from the existence of a complete classification of transport in the phenomenological framework of hydrodynamics. In \cite{Haehl:2014zda,Haehl:2015pja}, following a derivation of this eightfold classification of hydrodynamic transport compatible with the phenomenological axioms, we constructed an effective action for the adiabatic sector of hydrodynamics, which exhibits many features of our proposal in an ad hoc, but very explicit way.  In particular, it involves a doubling of fields, the dynamical degrees of freedom are Goldstone modes of broken symmetries, and finally the presence of a hidden $\UT$ symmetry is vital for constraining the allowed interactions to satisfy the second law. In this construction the entropy current can be explicitly seen as the $\UT$ Noether current. 

While the effective action presented in \cite{Haehl:2015pja} was not derived from first principles, it is worth recording here various issues encountered in attempts to construct effective actions for hydrodynamics. This program has seen attempts by several authors in recent years: \cite{Dubovsky:2011sj,Dubovsky:2011sk}, followed by \cite{Bhattacharya:2012zx,Haehl:2013hoa} among others, revived the analysis of effective actions for non-dissipative fluids which eschews the SK doubling. However, was shown in \cite{Haehl:2015pja} to capture only a part of hydrodynamic transport. On the other hand \cite{Grozdanov:2013dba,Grozdanov:2015nea} and  \cite{Kovtun:2014hpa,Harder:2015nxa} follow the SK prescription, but note various peculiarities present in such a framework, which were also highlighted in \cite{Haehl:2015pja}.\footnote{ Similar issues are encountered in recent attempts \cite{deBoer:2015ija,Crossley:2015tka} to derive such effective actions from a gravity following the paradigm outlined in \cite{Nickel:2010pr}.} We will simply note here that while various necessary ingredients are present in these analyses, ensuring that we capture all of hydrodynamic transport and developing a framework compatible with the microscopic QFT axioms leads us to the picture we espouse herein. 

It is also important in this context to emphasize the role of quantum anomalies in hydrodynamic transport. Indeed one of our first clues to the doubling of degrees of freedom came from attempts to write down effective actions for anomalous hydrodynamics \cite{Haehl:2013hoa}. The strategy here was to construct an off-shell action which upon restriction to hydrostatics reduces to the anomalous equilibrium partition function. Furthermore, attempts to write down a spacetime covariant hydrostatic partition function for theories with Lorentz anomalies, necessitated inclusion of 
the $\UT$ gauge field \cite{Jensen:2013rga}. Given the rigid structure imposed by anomalies it is rather reassuring that the ingredients we seek for the topological model are already present in the construction.

We take it as a justification of this manifesto that the various classes of fluid transport look a priori quite distinct, but find a natural and unifying explanation once the existence of doubling and $\UT$ is granted. In terms of computations, so far the strongest evidence for the correctness of our proposal lies in the fact that the above statements have been made precise in a mutually consistent and physically compelling way.

\subsection{Brownian branes: A topological gauge theory for entropy}

While it is useful to think about the  Mathai-Quillen (or MSR) formalism directly for hydrodynamics as alluded to above, it is in fact efficacious to go further and establish the theory of \emph{Brownian branes}. The idea here is to study probe branes of  an arbitrary dimension immersed into the fluid and which exhibit generalized Brownian motion as they are kicked around by the fluid. 

The dynamics of these Brownian branes can be described in terms by constructing appropriate world-volume sigma models which encode the effects of the fluid on this class of probes.  We will assume that the Brownian branes inherit the whole topological invariance of the fluid bulk. As a result the world-volume sigma models we seek are also topological. 

The Brownian brane topological sigma models have very much in common with the  topological sigma model describing the fluid itself. In fact, the world-volume theory of the space-filling Brownian brane coincides with that  of the fluid up to an overall prefactor.\footnote{  It is a common feature of large $N$ systems that the effective action of the whole system coincides with that of  the space-filling probe up to a prefactor \cite{Ferrari:2013aba}.}  We will find this to be  a very useful observation while shifting to the gravitational description in next section.

With the Brownian brane theory at hand, we can then go all the way down in dimensions until we get to the world-line theory of a Brownian particle. The corresponding topological sigma model is a twisted supersymmetric quantum mechanics which is well-known to be the world-volume description of Langevin theory for the  Brownian particle \cite{Das:1988vd,Birmingham:1991ty}. The Parisi-Sourlas supersymmetry \cite{Parisi:1979ka} of the Langevin theory  (and similar stochastic theories) then gets explained as a macroscopic consequence of the SK structure in the microscopic description. We review some of the salient features of this picture in Appendix \ref{sec:langevin}, drawing a clearer analogy between the MSR construction employed in traditional presentation of this material and the Mathai-Quillen formalism.

\subsection{Dissipation and entropy production}

Dissipation or the entropy production is the central (and arguably the most interesting) feature of a hydrodynamic effective field theory. This aspect will become more
important when we will shift to describing black holes where dissipation is at the heart of  information paradox.  It is thus natural to ask how dissipation
is described within the framework we have outlined in the last few subsections. We will begin by listing some crucial questions one would like the answers to 
\begin{enumerate}
\item In general, given a Schwinger-Keldysh effective theory, what  aspect of  it signals dissipation?
\item Given that dissipation often occurs only after a choice of boundary conditions (and a related `arrow of time'), how does one characterize such choices
of boundary conditions in a   Schwinger-Keldysh effective theory?
\item How does one understand in the Schwinger-Keldysh framework   the modern derivations of second law using Jarzynski relation \cite{Jarzynski:1997aa,Jarzynski:1997ab}?
\item In particular, in the context of fluid dynamical systems, how is a proposed $\UT$ gauge invariance related to entropy consistent with entropy production?\footnote{ One interesting possibility is that there is  a $\UT$ charge anomaly in the classical theory that is cured by quantum effects a la Green-Schwarz.}
\end{enumerate}
To these questions, a holographer would add a whole host of questions about black hole physics and information paradox. Each of these are central questions in 
non-equilibrium field theory and in this subsection we would like to describe our somewhat preliminary proposals on how they should be addressed. 

Consider our previous observation that unitarity implies that all Schwinger-Keldysh path integrals have a topological limit. One could ask whether the converse is true: does the existence of a topological limit guarantee unitarity? Since, dissipation is often associated with an effective non-unitary description for the relevant  degrees of freedom, one may naturally wonder if the existence of a  topological limit forbids dissipation. We will now argue that dissipation is still possible within the paradigm of Schwinger-Keldysh effective theories described in the previous section.  

By way of analogy, let us recall the classic proof of unitarity of gauge theories described in 
\cite{tHooft:1973pz}. The argument roughly has two ingredients: (i)  firstly, the largest time equation which is equivalent to the existence of the topological limit and, (ii) subsequently,  a proof that no ghosts appear in intermediate channels. Thus, even if the topological limit exists, one could have  non-unitary/dissipative dynamics  for the effective long distance modes, provided there are un-decoupled ghosts which mimic the physics of the microscopic degrees of freedom responsible for dissipation. One could think of this description as obtained by adding to the system under question a ghost system that replaces the actual
microscopic degrees of freedom and  serves as the entropy reservoir. It would be interesting to make this idea precise and check whether this is  a useful way to think about dissipation in effective theory.

We do have in Langevin system a toy model for dissipation with ghosts etc., where such an idea could be checked. One of the simplifying features of  Langevin theory is that it is a linear dissipative system. This enables one complete the construction of the theory efficiently and infer various physical consequences. For instance, the fluctuation-dissipation theorem, which is a macroscopic version of KMS relations, can be seen to naturally arise out of the underlying twisted supersymmetry.  In hydrodynamics the key difference is the non-linearity of dissipation, but the general idea of positivity of entropy production ought to arise from the underlying symmetry, providing thence a natural explanation for the second law in our proposal. It is then immediate to conjecture that the fluctuation relations and related features like Jarzynski relation \cite{Jarzynski:1997aa,Jarzynski:1997ab} for a Brownian brane  are also a consequence of the twisted supersymmetry. 

While the above statements are but a sketch of the plausible scenarios, the structural similarities between the Langevin dynamics and hydrodynamic effective actions constructed in \cite{Haehl:2014zda,Haehl:2015pja}  for the non-dissipative sector, lends us some confidence that the dissipative effective field theory can be constructed in a manner described above.

\section{Black hole story: Gravitational description}
\label{sec:gravity}

We finally  turn to the least conceptually understood description of fluids in terms of large AdS black holes, i.e., the fluid/gravity correspondence \cite{Bhattacharyya:2008jc}. Our job here is made difficult by the various confusions and problems that plague our understanding of quantum gravity  in the presence of horizons. Despite this we will now argue that when the fluid manifesto is taken seriously as a dual description of large AdS black holes,
one finds a remarkable concurrence of ideas that clarify hitherto confusing issues in gravity.

Let us recall the following four main ideas that have been forwarded as an explanation of how black hole physics should be understood.
\begin{enumerate}
\item The doubling structure of the field theory lies behind the construction of the black hole interior \cite{Papadodimas:2012aq,Papadodimas:2013wnh,Papadodimas:2015jra}. The paradigmatic example here is the thermofield double state, where left-right entanglement probes the black hole interior via AdS/CFT \cite{Maldacena:2001kr}.
\item Fluid dynamics can be formulated as a sigma model of fluid modes. In this picture the world volume fluid is dual to the so called membrane paradigm in black hole physics \cite{Nickel:2010pr}.\footnote{ The phrase `membrane paradigm' over the years has become a catchphrase for diverse  (and not all related) interpretations of horizon dynamics. We adhere here to the ideal interpretation: an effective description of the black hole interior in terms of surface dynamics.}
\item Black hole entropy is the Noether charge inside a black hole associated with the Killing horizon generator \cite{Iyer:1994ys}. We anticipate this Noether charge as being precisely the $\UT$ Noether charge, since the $\UT$ gauge transformations implement just thermal Lie-drags. 
\item Open strings ending on the stretched horizon of a black hole lead to a membrane paradigm explanation of black hole entropy \cite{Susskind:1993ws}.
\end{enumerate}
We would like to sketch a picture of what happens in the gravity description -- a picture that incorporates some version of the ideas above and the features described in the preceding few sections.

We conjecture that, for any dimension $p$ there is an object called $\UT$ $p$-brane in high temperature string theory  whose world volume theory is topological in the appropriate limit. As usual open strings end on this object thus leading to a  $\UT$ open string theory which is also topological. When they are placed inside AdS the geometry  ends at the   $\UT$ brane.\footnote{ This is analogous to an orientifold plane. In fact, we should probably think of $\UT$ brane as accompanied by a `SK orientifold plane', or $O_{SK}$ plane, to account for some of the unconventional CPT properties of the SK path integral. However, in the spirit of painting the broad brush outlines, we will ignore this subtlety in the following.} The geometry outside (and the closed  string theory that couples to $\UT$ open string theory) has an appropriate amount of twisted supersymmetry which  translates to topological invariance. In turn, this implies a SK quadrupling of all fields in AdS which is equivalent to  the usual un-quadrupled theory because of the BRST nature of twisted supersymmetry.

We can now invoke the open-closed duality for the $\UT$ branes. On the one hand the open string description leads to a construction of a fluid state in the CFT. On the other hand closed string dual is given by the $\UT$ brane geometry which we have associated to being the black hole state of AdS spacetime. This pair of statements then implies the fluid/gravity correspondence where the fluid state of CFT is dual to the black hole state of AdS.

The reader will recognize that the $\UT$ brane  is the long sought after brane of the membrane paradigm, with its adherent open string theory being the dual of black hole interior. We can further conjecture that $\UT$ string excitations are the quasinormal modes of the black hole with the $\UT$ open string tension being of the order of Hawking temperature.\footnote{ This should not be confused with $\UT$ brane tension which goes as inverse of Newton's constant.} At large temperatures, one can take an infinite tension limit and retain only the lowest quasinormal mode which is dual to fluid modes. The gauge field of $\UT$ then couples to the entropy current of the fluid realizing the idea that black hole entropy is a Noether charge. This picture thus incorporates all the ideas enumerated above and is dual to the emergence of fluid dynamics described in the last sections.

If the fluid manifesto is taken seriously as a dual description of large AdS black holes, such a picture is almost forced upon us. The doubling and the topological invariance of SK description are implemented in the twisted supersymmetry of the AdS bulk. The KMS invariance is imposed by the $\UT$ brane which signifies the emergence of $\UT$ gauge invariance in the IR. We can thus go further and conjecture that \emph{behind every
relativistic fluid, there is a  $\UT$  open string theory}. Equivalently, \emph{dual to every black hole interior  is a  $\UT$  open string theory}.

For example, we would like to identify the lower dimensional $\UT$ branes with the Brownian probe branes of CFT fluid dynamics on one hand and usual AdS probe branes on the other hand.  This is for example suggested by the manner in which we can use branes in AdS to probe the dynamics of Brownian particles \cite{deBoer:2008gu,Son:2009vu}.  This identification presumably gives a calculable way of completely fixing the $\UT$
open string theory.

We can  speculate on relating the  $\UT$  open string theory to other proposed ideas in black hole physics. Since $\UT$  open string leads to a gauge field that couples to entropy current, it is natural to assume that the ends of this open string carry entropy.\footnote{ Both of them should carry \emph{positive} entropy which probably is a consequence of SK orientifold.} It is natural to speculate that if there are two systems which are entangled and thus either of them carries entropy when the other is traced out, then they are endpoints of  a $\UT$  open string. An alert
reader would recognize here the idea of ER $=$ EPR \cite{Maldacena:2013xja} and the identification of the `quantum wormhole' with the $\UT$  open string. If this is right, then it follows that a classical ER wormhole is dual to a large number of  Euclidean $\UT$  open strings. We recognize here the famous feature of open-closed duality where the closed string exchange (here the ER wormhole) is dual to a  loop of stretched $\UT$  open strings.  We will thus propose that ER $=$ EPR is a direct consequence of the statement that $\UT$  open string theory and a closed string description in terms of black hole interior are dual descriptions.

The skeptical reader might rightly wonder where is  the twisted supersymmetry in the closed string description? The direct answer would require us to construct the computational framework and show that it agrees with classical gravity in the classical limit. But, at this stage, we would merely remind the reader of two
salient facts which support our proposal and provide circumstantial evidence that we are on the right track.
Firstly,  there is an equivalent of  replica trick in gravity \cite{Lewkowycz:2013nqa}. Secondly, in disordered systems, it is a well known statement that replica trick often can be recast in terms of twisted supersymmetry 
\cite{Efetov:1999aa,Kurchan:2002aa}. Neither of these statements are rigorous  enough yet to directly let us construct the twisted bulk theory capturing black hole dynamics, but we remain optimistic that the statements herein can be translated into a framework for developing SK supergravity.

Another tantalizing hint comes from Ooguri-Stromiger-Vafa (OSV) picture of black-hole partition function \cite{Ooguri:2004zv} whereby one writes down a doubled topological string theory which localizes to the Euclidean
partition function of certain asymptotically flat, supersymmetric black holes. The reader would discern  deep similarities between this picture and the Schwinger-Keldysh description that we espouse in this article. Perhaps a generalization of OSV picture to finite temperature AdS black holes would go a long way towards uncovering the Schwinger-Keldysh description on the gravity side.

Such a construction would also give us a concrete framework within which  questions about finite temperature black holes can be addressed. We dream of a future where this framework also allows us to tackle other vexing questions of  gravity with horizons both in relation to black holes and in cosmology.

\acknowledgments

It is a pleasure to thank Koushik Balasubramanian, Nabil Iqbal, Kristan Jensen, Veronika Hubeny, Juan Maldacena, Shiraz Minwalla, Suvrat Raju, Arnab Rudra, and Ashoke Sen for their insightful comments and perspicacious questions. FH wishes to thank Perimeter Institute for hospitality while this work was finalized. RL and MR would like to thank the organizers of Strings 2015, Bengaluru for giving an opportunity to present our work. RL would also like to thank 
International Center for Theoretical Sciences, Institute for Mathematical Sciences, Chennai Mathematical Institute and Tata Institute for Fundamental Research for hospitality during the course of this work.

FH is supported by a Durham Doctoral Fellowship and by a Visiting Graduate Fellowship of Perimeter Institute. Research at Perimeter Institute is supported by the Government of Canada through Industry Canada and by the Province of Ontario through the Ministry of Research and Innovation. RL gratefully acknowledges support from Institute for Advanced Study, Princeton. MR was supported by the European Research Council under the European Union's Seventh Framework Programme (FP7/2007-2013), ERC Consolidator Grant Agreement ERC-2013-CoG-615443: SPiN (Symmetry Principles in Nature) and by the  FQXi  grant "Measures of Holographic Information"  (FQXi-RFP3-1334).

\appendix

\section{Example: Langevin theory}
\label{sec:langevin}

We will now consider a simple system which serves as a paradigmatic example for the ideas presented herein.
We focus on the well studied Langevin theory  which describes Brownian motion of a particle in a thermal medium.
Our logic will be to assume the symmetries of the SK construction noted in the main text and demonstrate that the resulting dynamics is precisely one captured by the Langevin equation. We should note at the outset that the connection between stochastic systems of the Langevin type and topological models has been known for a long time \cite{Parisi:1979ka} (see also \cite{Das:1988vd},  \cite{Birmingham:1991ty}, and the textbook \cite{ZinnJustin:2002ru}), but phrasing it in our language allows us to demonstrate the general principles.

Consider a Brownian particle which is characterized by a single degree of freedom, its position $x(t)$. We will assume that this particle is subject to a time independent (conservative) force, arising from a potential $U(x)$, in addition to the friction it encounters from the fluid medium it is immersed in. The stochastic Brownian motion is then described by the Langevin equation:
\begin{equation}
m\frac{d^2x}{dt^2} +\frac{\partial U}{\partial x}  +  \nu\ \delKMS x
= \mathbb{N} \,,
\label{eq:langevin}
\end{equation}
where we have normalized various terms with later applications in mind. The following comments concerning the Langevin equation as a toy model for hydrodynamics are in order:
\begin{enumerate}
\item The kinetic term for the particle  gives us a part of dynamical response
that is adiabatic and  belongs to Class $\LS$ in the  terminology of \cite{Haehl:2015pja}.
\item The second term denotes the static response to the background potential $U(x)$  and hence is Class $\PS$.
\item The third term then incorporates dissipation by introducing a viscous drag parametrized by the coefficient $\nu$. This is Class D transport.
\item The term on the right hand side finally adds a  stochastic noise $\mathbb{N}$ which we assume is independent of $x$. This did not enter into the considerations of \cite{Haehl:2015pja} as the noise is related to presence of dissipation via the fluctuation-dissipation theorem. In writing an effective action we will require that the noise is always integrated out. The noise term is assumed to be drawn from a Gaussian ensemble (as appropriate for a linear dissipative system), which we specify explicitly below, see \eqref{eq:LangMeasure}. 
\end{enumerate}

\subsection{Twisted supercharges for the  Langevin system}
\label{sec:QsL}

By the logic espoused in the main text, the presence of the dissipative term requires that we formulate the problem in terms of a SK path integral. Let us first do so by identifying the degrees of freedom that make up the SK quadruplet. To do so, we first introduce a doubling of the Brownian particle's position, by $
x \to \{x_\skR, x_\skL\}$. It will be useful to a-priori realize that the Langevin description is naturally adapted to the advanced/retarded basis of fields (as opposed to the $\{\skR,\skL\}$ Keldysh basis), cf., footnote \ref{fn:advret}. Introduce therefore
the \emph{retarded} field  $x(t)$ and its  \emph{advanced} field partner $f_\psi(t)$ via:\footnote{ The Brownian particle is assumed to be bosonic  and so $\FSgn{} =1$.}
\begin{equation}
x \equiv -i \delKMS^{-1} \, (x_\skR - e^{-i\,\diffBi} \, x_\skL ) 
 \,, \qquad 
 f_\psi \equiv x_\skR- x_\skL
\label{eq:arL}
\end{equation}	
The definition above can be treated as a classical differential equation for the retarded field by properly inverting the operator $\delKMS$ subject to the physical initial conditions. We recall that the operator $\delKMS$  measures the deviation from the KMS condition and thus controls the dissipative terms. It involves thermal time translations (via $\diffBi$). In the high temperature limit  relevant for the current discussion, we can approximate it as
\begin{equation}
\delKMS\equiv - i \left( 1 - e^{-i\diffBi}\right) \approx \diffBi \equiv  \Kbeta_0 \frac{d}{dt} \,.
\end{equation}
With this definition it is then clear that $x(t)$ can be obtained by integrating up the Keldysh basis of fields. 

Following the discussion of \S\ref{sec:micro} we will embed the position of the Brownian particle into an SK quadruplet $\{x,\psi,\bar{\psi},f_\psi\}$. The action of SK supercharges follows from \eqref{eq:QSKdefRL}:
\begin{equation}\label{eq:QSKLangevin}
\begin{split}
\gradcomm{\QSK}{x}&=  \psi,\quad
\gradcomm{\QSK}{\psi } = 0\, ,\quad
\gradcomm{\QSK}{{\bar \psi}} = - f_\psi\, ,\quad
\gradcomm{\QSK}{f_\psi} = 0\,, \\
\gradcomm{\QSKb}{x} &= {\bar \psi}\,,\quad
\gradcomm{\QSKb}{{\bar \psi} } = 0 \,,\quad
\gradcomm{\QSKb}{\psi} =  f_\psi \,,\quad
\gradcomm{\QSKb}{f_\psi} = 0 \,.
\end{split}
\end{equation}
The KMS supercharges act on this basic multiplet as
\begin{equation}\label{eq:QKMSLangevin}
\begin{split}
\gradcomm{\QKMS}{x}&= 0,\quad
\gradcomm{\QKMS}{\psi } = 0\, ,\quad
\gradcomm{\QKMS}{{\bar \psi}} =- i\delKMS x\, ,\quad
\gradcomm{\QKMS}{f_\psi} = -i\delKMS \psi\,, \\
\gradcomm{\QKMSb}{x} &= 0\,,\quad
\gradcomm{\QKMSb}{{\bar \psi} } = 0 \,,\quad
\gradcomm{\QKMSb}{\psi} = i\delKMS x \,,\quad
\gradcomm{\QKMSb}{f_\psi} = -i\delKMS \bar{\psi} \,.
\end{split}
\end{equation}
Finally, we have the bosonic generator
\begin{equation}\label{eq:Q0Langevin}
\begin{split}
\gradcomm{\Qzero}{x}&= 0,\quad
\gradcomm{\Qzero}{\psi } = 0\, ,\quad
\gradcomm{\Qzero}{{\bar \psi}} = 0\, ,\quad
\gradcomm{\Qzero}{f_\psi} = -i\delKMS x\,.
\end{split}
\end{equation}
The four non-local charges $\{ \Qzero,\QKMS, \QKMSb,\,i\delKMS\}$ form a Schwinger-Keldysh quartet of thermal (super-)translations:
\begin{equation}
\gradcomm{\QSK}{\Qzero} = \QKMS \,,\quad \gradcomm{\QSKb}{\Qzero} = \QKMSb \,, \quad \gradcomm{\QSK}{\QKMSb} = - \gradcomm{\QSKb}{\QKMS} = i \delKMS \,,
\end{equation}
or as a diagram
\begin{equation}
\begin{aligned}
\begin{tikzcd}
&\quad \Qzero \quad \arrow{ld}{\!\!\!\QSK} \arrow{rd}[below]{\!\!\!\!\QSKb}  &   \\
\QKMS \arrow{rd}{\!\!\! -\QSKb \quad\;\;\;\;\;\;\QSK} & & \QKMSb \arrow{ld}\\
&   i \delKMS &
\end{tikzcd}
\end{aligned}
\label{eq:nt2lang}
\end{equation}

These algebras can be inferred from \eqref{eq:QSKdefRL} and \eqref{eq:QKMSdefRL} by performing the basis rotation from the Keldysh basis to the advanced/retarded basis.

The algebra of the charges $\{\QSK, \QSKb, \QKMS, \QKMSb, \Qzero, i \delKMS\}$ which is encoded in 
\eqref{eq:nt2lang} is precisely the ${\cal N}_T =2$ extended equivariant algebra of \cite{Dijkgraaf:1996tz}. This will be made manifest when we discuss the equivariant construction in Appendix \ref{sec:LangevinCartan} below.

\subsection{The MSR effective action for Langevin dynamics}  
\label{sec:LangevinMSR}

Let us now see the emergence of the quadruplet $\{x,\psi,\psib, f_\psi\}$ from an effective action perspective. We will first construct the effective action following a standard discussion in stochastic quantization and verify that the resulting action has the twisted supercharges constructed above.

The MSR construction \cite{Martin:1973zz} (see also \cite{Janssen:1976fk,DeDominicis:1977fw} and the Parisi-Sourlas construction \cite{Parisi:1979ka})  provides a prescription for rewriting this stochastic system in SK language. The basic point is the following: we want to study correlation functions after ensemble averaging the correlators. Since we want to ensemble average the correlators,
we have to normalize the path-integral to unity (i.e., pass to a SK description).
We begin by considering the space  of solutions for a particular noise realization and writing a normalized measure
which integrates to unity:
\begin{equation}\label{eq:measureLangevin}
\begin{split}
[dx]\ \delta (E_x + \mathbb{N} )\ \det\prn{\frac{\delta E_x}{\delta x} }\,,
\end{split}
\end{equation}
where we have defined the non-noise part of Langevin equation to be the negative of the left hand side of \eqref{eq:langevin} for brevity, viz.,
 \begin{equation}
 \begin{split}
 E_x \equiv  -m\frac{d^2x}{dt^2}-\frac{\partial U}{\partial x} - \nu\,  \delKMS x \,.
 \end{split}
 \end{equation}
From this point of view, it is now very natural to introduce the advanced Lagrange multiplier field $f_\psi$ to exponentiate the delta-function, and a pair of Grassmann valued scalar ghosts $\{ \psi, \bar \psi \}$ to write the functional determinant using the standard Faddeev-Popov trick. If we further take a Gaussian noise realization, we can then rewrite the measure \eqref{eq:measureLangevin} including the Gaussian weight for $\mathbb{N}$ as
\begin{equation}\label{eq:LangMeasure}
\begin{split}
[dx]\int [df_\psi] [d\bar{\psi}] [d\psi]\ \exp\prn{i\int dt\ \brk{
f_\psi\ E_x +f_\psi \,\ \mathbb{N}
+ \bar\psi\prn{\frac{\delta E_x}{\delta x} } \psi
+ i\,\frac{\mathbb{N}^2}{4\nu}\, 
} } \,. \\
\end{split}
\end{equation}
Here we have fixed the width of the Gaussian noise using fluctuation-dissipation theorem (or alternately Einstein-Smoluchowski relation).

Integrating out the noise $\mathbb{N}$, we can now obtain the generating function for the Brownian correlators, i.e.,
\begin{equation}
\begin{split}
[dx]\int [df_\psi] [d\bar{\psi}] [d\psi]\ \exp\prn{i\int dt\ \brk{
f_\psi\ E_x + i \, \nu\,f_\psi^2
+ \bar\psi\prn{\frac{\delta E_x}{\delta x} } \psi
} } \,. \\
\end{split}
\end{equation}
Thus, the SK Lagrangian for the Langevin system is given by
\begin{equation}
\begin{split}
\Lag_{SK} &=
f_\psi\ E_x + i \, \nu \, f_\psi^2
+ \bar \psi\prn{\frac{\delta E_x}{\delta x} } \psi\\
&= -\brk{f_\psi\ \frac{\partial U}{\partial x} +  \bar\psi \frac{\partial^2 U}{\partial x^2}\psi }
- m \brk{ f_\psi\ \frac{d^2x}{dt^2} + \bar\psi\ \frac{d^2\psi}{dt^2} }\\
&\qquad -\nu \brk{f_\psi\  \delKMS x -\bar\psi\  \delKMS \psi} + i \, \nu \, f_\psi^2\,.
\end{split}
\label{eq:LSKLan}
\end{equation}
As promised the single Brownian field $x$ is now part of a quadruplet $\{x,\psi,\bar{\psi},f_\psi\}$, filling out nicely the multiplet structure anticipated in \S\ref{sec:micro} and confirms the expectations of Appendix \ref{sec:QsL}. It is easy to check explicitly that the supercharge actions given in the latter subsection indeed act as advertised and leave $\Lag_{SK}$ invariant. We will do this below explicitly, but in the process will take the opportunity to explain the construction in the mathematical framework of equivariant cohomology appropriate for this class of topological field theories.

\subsection{Equivariant description}
\label{sec:LangevinCartan}

Let us now recover the Langevin Lagrangian \eqref{eq:LSKLan} from an equivariant cohomology construction. The notion of equivariant cohomology is relevant in circumstances where we have an action of a group on a topological space and we wish to define cohomological observables compatible with the group action. In the present circumstance we will view the group action as being generated by the (abelian) group of thermal time translations acting on the space of field configurations. Basically we want to quotient the configuration space subject to the Langevin equation of motion such that we ensure that Euclidean time translations remain a symmetry and impose the KMS condition, which in turn leads to the correct fluctuation-dissipation theorem.

There are two equivalent models for the equivariant cohomology construction called the Weil and Cartan models. In both cases we start with a gauge field which encodes the group action on the fields; the underlying supersymmetry requires that the gauge field be part of a quadruplet of fields (just like the position multiplet for $x$). In addition to ensure that we have a gauge covariant action we have to introduce corresponding 
Faddeev-Popov ghosts and anti-ghosts; but these too have to be upgraded into corresponding multiplets 
to make all the symmetries manifest. In the Weil model therefore we have 12 fields forming a suitable  dodecuplet. The various fields should be viewed as ghosts, ghosts for ghosts (to fix residual gauge symmetry) etc., and can be efficiently represented in a superspace formalism \cite{Horne:1988yn}. If we consider a gauge fixed problem where we set the Faddeev-Popov ghosts to zero, we are left with 9 fields
which comprise of the data in the Cartan model. Since we are dealing with a one-dimensional theory, the gauge fields are also easily gauge fixed; we can therefore set the entire gauge multiplet to zero leaving behind a set of five fields which correspond to the  multiplet of gauge parameters. We denote these as $\{\phi, \eta,\phi_0,\bar{\eta},\bar{\phi}\}$ with ghost numbers $(2,1,0,-1,-2)$ respectively.

Our motivation for introducing another set of ghostly degrees of freedom has been very sketchy, but the essential rationale is that is allows us to exploit the standard construction of balanced topological field theories of \cite{Dijkgraaf:1996tz}. In particular, the Cartan charge we define below allows us to write down the Lagrangian for the Langevin system by examining simply the set of cohomologically exact contributions to an effective action as desired for a topological field theory. We will give a more comprehensive discussion of these statements in a forthcoming publication \cite{Haehl:2015ab}.

Including these additional gauge parameter multiplet ghost fields $\{\phi, \eta,\phi_0,\bar{\eta},\bar{\phi}\}$  we can construct the action of the Cartan supercharges on the basic ``position" multiplet $\{x,\psi,\bar{\psi},f_\psi\}$. They act as 
\begin{equation}\label{eq:PositionCartan}
\begin{split}
  \gradcomm{\QC}{x} = \psi  \,,\qquad& \gradcomm{\QC}{\bar\psi} = - f_\psi + \phi_0 \delKMS x \,,\\
  \gradcomm{\QC}{\psi} = \phi \delKMS x \,,\qquad& \gradcomm{\QC}{f_\psi} = \phi_0 \delKMS\psi - \phi \delKMS\bar{\psi} +\eta \delKMS x  \,,\\
  \gradcomm{\QCb}{x} = \bar\psi \,,\qquad& \gradcomm{\QCb}{\bar\psi} =  \bar\phi \delKMS x  \,,\\
  \gradcomm{\QCb}{\psi} =f_\psi  \,,\qquad& \gradcomm{\QCb}{f_\psi} =  \bar\phi \delKMS \psi \,,\\
\end{split}
\end{equation}

Note that these relations suggest a natural identification of the equivariant charges with the Schwinger-Keldysh charges of \eqref{eq:QSKLangevin} after gauge fixing the gauge parameter sector to zero, $\{\phi, \eta,\phi_0,\bar{\eta},\bar{\phi}\} = 0$. 
We refrain from performing this gauge fixing and suggest an even closer relation between the microscopic charges $\{\QSK,\QSKb,\QKMS,\QKMSb,\Qzero\}$ and the Cartan charges of $\{\Q,\Qb\}$ the low-energy theory via
\begin{equation}
\begin{split}
  \QC = \QSK+ i \phi_0 \QKMS - i \phi \,\QKMSb + i \eta \,\Qzero \,, \qquad \QCb = \QSKb + i \bar \phi  \QKMS\,.
\end{split}
\end{equation} 
Note that the asymmetry between $\QC$ and $\QCb$ is not a fundamental feature, but simply due to our choice of convention for the distribution of $\eta$ and $\phi_0$ in \eqref{eq:PositionCartan}. It would be straightforward to formulate \eqref{eq:PositionCartan} in a symmetric fashion.\footnote{ As argued in \cite{Dijkgraaf:1996tz} and emphasized in \cite{Zucchini:1998rz} the extended equivariant cohomological algebra has a  $\mathfrak{sl}_2$ automorphism symmetry under which $\{\QKMS,\QKMSb, \Qzero\}$ transform as a triple.}

Finally, we need to define the action of the supercharges on the gauge parameter multiplet:
\begin{equation}\label{eq:GaugeCartan}
\begin{split}
  \gradcomm{\QC}{\phi_0} = \eta   \,,\qquad& \gradcomm{\QCb}{\phi_0} = \bar{\eta}  \,,\\
  \gradcomm{\QC}{\phi} = 0\,,\qquad& \gradcomm{\QCb}{\phi} = -\eta \,,\\
  \gradcomm{\QC}{\bar \phi} = -\bar \eta  \,,\qquad& \gradcomm{\QCb}{\bar \phi} =0 \,,\\
  \gradcomm{\QC}{\eta} = (\phi,\phi_0)_\Kbeta \,,\qquad&
  \gradcomm{\QCb}{\eta} = (\phi,\bar \phi)_\Kbeta\,,\\
  \gradcomm{\QC}{\bar\eta} = (\bar \phi,\phi)_\Kbeta\,,\qquad&
  \gradcomm{\QCb}{\bar\eta} = (\bar \phi,\phi_0)_\Kbeta  \,,
\end{split}
\end{equation}
where $(\hat{\Op{A}},\hat{\Op{B}})_\Kbeta \equiv \hat{\Op{A}}\delKMS \hat{\Op{B}} - (-)^{\Op{A}\Op{B}} \,\hat{\Op{B}} \delKMS \hat{\Op{A}}$.\footnote{ Here, $(-)^{\Op{A}\Op{B}}$ is $1$ if either of $\hat{\Op{A}}$ and $\hat{\Op{B}}$ is Grassmann even, and $-1$ otherwise.} 

We remark that the above construction ensures that the Cartan charges $\QC$, $\QCb$ square to $\UT$ gauge transformations with gauge parameters $\phi$ and $\bar{\phi}$, respectively. Further, they anticommute to a $\UT$ gauge transformation with gauge parameter $\phi_0$.

Now turning to the Langevin system, we can construct the various pieces of \eqref{eq:LSKLan} in a balanced cohomological way, i.e., as $\Qb \Q$-exact terms. The three pieces of \eqref{eq:LSKLan} are constructed, respectively, as follows:
\begin{itemize}
\item {\bf Potential term:} Consider the following equivariant term:
\begin{equation}\label{eq:LangevinQQ1cartan}
\begin{split}
  \gradcomm{\QCb}{\gradcomm{\QC}{-U(x)}} &=  -\left(f_\psi \, \frac{\partial U}{\partial x} + \bar \psi \, \frac{\partial^2 U}{\partial x^2} \, \psi  \right)\,.
\end{split}
\end{equation}
This is just the potential term of the Langevin Lagrangian.

\item {\bf Mass term:} The mass term in \eqref{eq:LSKLan} works essentially the same way as the potential term. We can generate it as
\begin{equation}\label{eq:LangevinQQ2cartan}
 \gradcomm{\QCb}{\gradcomm{\QC}{\,\frac{m}{2} \, (\delKMS x)^2}} = -m\left( f_\psi \delKMS\delKMS x + \bar \psi \delKMS\delKMS\psi \right)+ \delKMS(\ldots) \,,
\end{equation}
where the last term denotes total derivatives, which we can drop in the Lagrangian.

\item {\bf Friction term:} Finally, to obtain the friction term, we consider the following balanced combination
\begin{equation}\label{eq:LangevinQQ3cartan}
\begin{split}
  \gradcomm{\QCb}{\gradcomm{\QC}{-i\,\nu\, \bar{\psi}\psi}} &= -i\,\nu\,\big(-f_\psi^2 +\bar{\phi} \psi \delKMS\psi +\phi_0 \, f_\psi \delKMS x - \phi\bar{\phi} (\delKMS x)^2 \\
  & \qquad\quad\;\;\; + (\bar \eta \psi + \eta \bar \psi) \delKMS x - (\phi_0 \psi - \phi \bar{\psi}) \delKMS \bar{\psi}\big)  \,.
\end{split}
\end{equation}
\end{itemize}
We should note here that the set of terms written above exhaust the possibilities for cohomologically trivial terms as can be inferred from a result proved in \cite{Dijkgraaf:1996tz} (see Theorem 2.1 and the discussion of section 4.3 there). What we have written down here is nothing but the equivariant construction of a special case of the twisted supersymmetric quantum mechanics theory constructed in \cite{Witten:1982im} originally.

\paragraph{Gauge fixing:} Let us now make contact with the earlier discussion of a Langevin effective action using the MSR formalism.  To identify with the result in Eq.~\eqref{eq:LSKLan} we perform a particular gauge fixing by examining the structure of the friction terms.  Consider setting
\begin{equation}
\phi = \eta = \bar \eta = \bar \phi = 0\,,\quad  \phi_0 = -i\,.
\end{equation}
After applying this gauge fixing of the gauge parameter quintuplet, we get the friction term in \eqref{eq:LangevinQQ3cartan} to reduce to (up to total derivatives)
\begin{equation}
\begin{split}
  \gradcomm{\QCb}{\gradcomm{\QC}{-i\,\nu\, \bar{\psi}\psi}} &=   i\,\nu \, f_\psi^2-\nu \,(  f_\psi \delKMS x  - \bar \psi \delKMS {\psi})   + \delKMS(\ldots)\,,
\end{split}
\end{equation}
where we recognize the friction term and a Gaussian noise contribution of the Langevin Lagrangian \eqref{eq:LSKLan}. As an added bonus we see the cohomological origins of the fluctuation-dissipation relation: the coefficients in front of the kinetic term for the ghosts $\psi, \bar{\psi}$ are correlated with that in front of the Lagrange multiplier field $f_\psi$ as desired.

To summarize, the Langevin Lagrangian \eqref{eq:LSKLan} can be written equivariantly as
\begin{equation}
\mathcal{L}_{SK} = \gradcomm{\Qb}{\gradcomm{\Q}{\,\frac{m}{2} \, (\delKMS x)^2 -U(x) - i \,\nu\,\psib \psi }}\bigg{|}_{\phi = \eta = \bar \eta = \bar \phi = 0\,,\,  \phi_0 = -i} 
\end{equation}
This concludes our derivation of the Langevin Lagrangian in the language of equivariant cohomology. We have seen how equivariant supercharges naturally emerge in this construction and that they can be related to the SK and KMS algebras in a simple way. We postpone a more detailed exposition of the topic and an equivalent description of this construction in the so-called Weil model to a future publication \cite{Haehl:2015ab}.

\section{Should $\UT$ be gauged in the field theory? }
\label{sec:utgauge}

In this appendix, we will more closely examine our proposal for a gauge theory underlying entropy current.  In particular, the question which will concern us revolves around whether $\UT$ should 
be considered a global symmetry or should we consider it as a gauge redundancy.\footnote{ We thank Juan Maldacena, Shiraz Minwalla and Spenta Wadia for discussions on this point.}

Let us begin by noting the following peculiar features of  $\UT$:
\begin{enumerate}
\item Since $\UT$ acts on many of the fields by thermal translation, the symmetry itself depends 
on the local temperature of the state under question. It is thus a \emph{state-dependent} symmetry.
\item More specifically, it is a state-dependent time translation with entropy as its charge (which should be distinguished from state-independent time  translation with energy as its charge).
\item In particular, this means that when we apply a perturbation to the fluid which changes the 
temperature of the fluid, we should continuously redefine what the symmetry is from one instant of time to the next.
\item Another peculiar feature of the kind of theories we describe in this article is that at least as 
macroscopic effective theories they need not be unitary by themselves -- and thus ghosts need not necessarily decouple and in fact may play the physical role as a proxy for the microscopic degrees
of freedom responsible for dissipation.
\end{enumerate}
These features suggest that $\UT$ as a symmetry and its gauging can involve essentially novel features and in general, we should be careful in extending our usual intuitions to the framework under study. We will leave a detailed study of these interesting features to future work. Our purpose here is to merely set up how these peculiar features might be germane to fundamental questions about how
 $\UT$  symmetry acts. 

For example, one of the basic problems with positing a symmetry behind entropy current is the problem of non-conservation of the entropy current. Does this mean that $\UT$  symmetry  is necessarily broken or is it just an indication that $\UT$ is a state-dependent symmetry which is continuously redefined with time which admits specific forms of non-conservation? Note that while entropy is not conserved, its non-conservation is heavily constrained by the second law. If   $\UT$  symmetry is indeed broken, it is difficult to see why a generic breaking should be second law consistent. A further question concerns the role played by un-decoupled ghosts in the process of entropy production and how this modifies our usual intuitions about symmetries, currents and conservation.

In this article, we have taken the viewpoint that $\UT$  symmetry should be thought of as being gauged. We will present now three main arguments why we think this is the right way to think about $\UT$.

First, remember that the origin of $\UT$ ultimately lies in the (anti-)periodicity conditions on fields 
imposed in the Euclidean description. The thermal translations are \emph{gauged} in this description in the sense that field values separated by thermal translations are \emph{identified}. It is then plausible to believe that after the analytic continuation to real time and a macroscopic limit, Euclidean
periodicity survives as a local gauge invariance. This argument can be made precise if we directly work in the long distance limit of Euclidean description which becomes the hydrostatic limit of fluid dynamics. The hydrostatic partition function is known to be naturally defined on the spatial manifold which in Euclidean description is obtained via the Kaluza-Klein reduction on the thermal circle. After 
analytic continuation to real time, there is no thermal circle anymore but the same spatial manifold can be obtained by modding the space-time with thermal translations. This modding out is the reason why equivariant connections and curvatures play an important role in the description of anomalies 
in hydrostatics. We take these combination of facts as a strong suggestion that $\UT$  symmetry 
should be treated as being gauged.

Our second argument concerns the state-dependence of $\UT$ as a symmetry and the state-dependence in the general Schwinger-Keldysh description because of the introduction of  $\UT$.
If one treats this state-dependence as an artifact of our description, then it is natural to expect that state-dependence in some sense should be gauged. Gauging  $\UT$, which is the major source of initial state-dependence in Schwinger-Keldysh description, might thus be a way to show that the state-dependence can be gauged away. Given our present understanding, this is 
indeed a weak argument  but it is suggestive.

Our third argument uses holography. If $\UT$ is indeed a global emergent symmetry in thermal systems, the rules of AdS/CFT suggests that there be an emergent gauge field in the bulk corresponding to the global symmetry in the boundary. Such a suggestion however faces a series of theoretical and empirical issues well-known in the context of  `entropic force' proposals in gravity.
Gauging  $\UT$ seems to be the most straightforward way to sidestep these issues.

These indirect arguments  suggest that  treating $\UT$  as being gauged is the right thing to do.
As to whether these arguments stand up to detailed scrutiny, we  leave  to the  future.


\providecommand{\href}[2]{#2}\begingroup\raggedright\endgroup

\end{document}